\documentstyle[12pt,aasms4]{article}

\newcommand{\etal}{$et \; al.$~}
\newcommand{\elec}{$e^- \,$}

\def\lesssim{\mathrel{\hbox{\rlap{\hbox{%
 \lower4pt\hbox{$\sim$}}}\hbox{$<$}}}}
\def\gtrsim{\mathrel{\hbox{\rlap{\hbox{%
 \lower4pt\hbox{$\sim$}}}\hbox{$>$}}}}

\def\arcmin{\hbox{$^\prime \,$}}
\def\arcsec{\hbox{$^{\prime\prime \, }$}}

\def\farcs{\hbox{$.\!\!^{\prime\prime}$}}

\def\micron{\hbox{$\mu$m}}

\begin{document}

\title{Variable Galaxies in the Hubble Deep Field}
\author{Vicki L. Sarajedini}
\affil{Wesleyan University, Middletown, CT 06459}
\authoremail{vicki@astro.wesleyan.edu}
\author{Ronald L. Gilliland}
\affil{Space Telescope Science Institute, 3700 San Martin
Drive, Baltimore, MD 21218}
\authoremail{gillil@stsci.edu}
\author{M. M. Phillips}
\affil{Carnegie Institution of Washington, Las Campanas Observatory,
Casilla 601, La Serena, Chile}
\authoremail{mmp@ociw.edu}

\begin{abstract}

We present results from a study to detect variable galaxies in
the Hubble Deep Field North.  The goal of this project is to
investigate the number density of AGN at z$\simeq$1 through
the detection of variable galaxy nuclei.
The advantage of HST is the ability to do
accurate photometry within smaller apertures, thus allowing us
to probe much lower AGN/host galaxy luminosity ratios than can be done
from the ground.

The primary data sets analyzed for galactic variability follow from the original
HDF-N observations (Williams \etal 1996) in December 1995 and a second epoch
obtained two years later (Gilliland, Nugent \& Phillips 1999).
The second epoch data consists of 36 exposures in F814W with a total integration
time of 63000 seconds (compared to 58 exposures and a total of 123600 seconds in
the original HDF-N).

We have detected nuclear variability at or above the 3$\sigma$ level in 8
of 633 HDF galaxies at I$_{814}$$\lesssim$27.  
Only 2 detections would be expected by chance in a normal distribution.
At least one of these 8 has
been spectroscopically confirmed as a Seyfert 1 galaxy.  
Based on the AGN structure function for variability, the estimated luminosity
of the varying component in each galaxy lies in the range
-19.5$\lesssim$M$_B$$\lesssim$-15.0.
We construct an upper limit to the luminosity function for the variable nuclei and
compare this to the local Seyfert LF and the LF for QSOs at z$\simeq$1.
Assuming we have detected all Seyfert-like nuclei in the HDF-N, 
we find no evidence for an increase in the number density of AGN
at M$_B$$\simeq$-19 (H$_o$=75 km/s/Mpc,
q$_o$=0.5).  From this study, we estimate that 
$\sim$1--3\% of field galaxies with
I$\lesssim$27 may contain a nuclear AGN.

\end{abstract}


\section{Introduction}

Variability is a well known property of active galaxies.
A high incidence of variability has been found among QSO samples (Bonoli \etal
1979; Kron \& Chiu 1981; Hawkins 1986).  Koo, Kron \& Cudworth (1986) showed
that $\sim$80\% of their spectroscopic and color selected quasars in 
Selected Area 57 (SA57) were variable over the time period of 11 years.
In addition to quasars, lower luminosity AGNs such as Seyfert 1s and 
even Seyfert 2s and
LINERs have been shown to exhibit both line and continuum variability
(see Aretxaga, 1997 and references therein).
Bershady \etal (1998) detected 14 extended variable
objects in SA57 with Seyfert-like spectral characteristics.  The
variability amplitude was generally higher for AGN of lower luminosity
making this technique well suited for the selection of intrinsically faint
QSOs and Seyfert galaxies.

A prominent gap in our current understanding of the AGN phenomenon is the
connection between high luminosity QSOs and AGN of lower luminosity.
In particular, knowledge of the shape and slope of the low luminosity
end of the AGN luminosity function (LF) over a range of redshifts is critical 
in differentiating AGN
evolution models such as pure luminosity, pure density, and luminosity 
dependent density evolution. 
Locally, Huchra \& Burg (1992) have constructed the LF
for spectroscopically selected Seyfert galaxies, but only brighter than
M$_B$$\simeq$-18.5, while Ho \etal (1997) have detected fainter nuclei
in local galaxies but lack the photometric information to construct the
LF.  At higher redshifts, however, spectroscopically selected samples as well
as most other AGN selection techniques are quickly limited to active nuclei
which dominate the host galaxy light and are thus likely to be intrinsically
brighter (e.g. Hartwick \& Schade 1990).  
X-ray selection has produced samples of higher redshift, lower luminosity 
AGN, but without precise optical-to-X-ray flux ratios, these cannot be easily
compared with the multitude of optical data for high luminosity QSOs.
Through the optical detection of faint,
variable nuclei within
brighter host galaxies, low-luminosity
AGN (LLAGN) can be selected for the purpose of studying 
the relationship between these objects
and higher luminosity QSOs at all redshifts.

In this study, we use HST images of the Hubble Deep Field North separated by
2 years to search for variable objects with emphasis on investigating
LLAGN at redshifts of z$\simeq$1.   The advantage of HST is the ability
to do accurate photometry within smaller apertures than can be done
from the ground, allowing us to probe much lower AGN/host galaxy luminosity 
ratios.

The primary data sets analyzed for galactic variability follow from the original
HDF-N observations (see Williams \etal 1996) in December 1995 and a second epoch
obtained two years later (Gilliland, Nugent \& Phillips 1999, hereafter: GNP).
We will only discuss data from the F300W and F814W filters for which extensive 
observations are
available in both epochs (F606W will be brought in as a special case).
The second epoch data consists of 36 exposures in F814W with a total integration
time of 63000 seconds (compared to 58 exposures and a total of 123600 seconds in
the original HDF-N).  Also available from the second epoch are 18 exposures
in F300W with a total integration time of 27300 seconds (compared to 95
exposures and a total of 178800 seconds in the original).
Observations in both epochs were taken using the same orientation (to within a 
9\arcmin
roll difference) and pointing (to $\sim$1\arcsec ) on the sky, and each was
similarly dithered.
Sky background levels were also very similar for the two epochs separated by
731.9 days, or 2.00 years.
Critical sampling for the inherently under-sampled WF images is provided by
the substantial sub-pixel dithering.
The repeated exposures and use of multi-pixel dither offsets in each epoch
allow for excellent rejection of both hot pixels and cosmic rays.

We describe the observations and details concerning the
galaxy photometry in \S 2.  
A description of each variable
galaxy AGN candidate detected in the HDF-N is contained in \S 3.  
We then compute the luminosity function of variable nuclei 
in the HDF-N in \S 4 and compare it to the local number 
densities of AGN.  In \S 5 we present a summary of this work and 
our conclusions.  

\section{Photometry}

\subsection{Creation of Over-sampled, Summed Images}

A thorough discussion of the data reductions is given in GNP.
For each epoch and each bandpass all available frames are combined to produce
a single clean image.
Hot pixels and cosmic rays are identified and excluded from contributing to
the combined image.
The individual exposures are taken with different exposure times
and varying levels of background light.  The combined image production
utilizes weights based on readout noise, Poisson statistics
on the background and source counts to provide a minimal noise in the
final combined, over-sampled image.
The adopted algorithm for creation of an over-sampled image uses frame-by-frame
registration offsets accurate to $\sim$ 0.02 pixels for F814W and $\sim$ 0.1
for F300W and an accurate model for geometric distortion to maintain optimal
resolution.
Scattered light patterns affecting a small subset of the images are removed
before combining.

To facilitate easy qualitative and quantitative searches for change between
the two epochs, the second epoch images are constructed to align with an
external galaxy defined reference grid from the first epoch.
The resulting alignments are good to {\em maximum} errors of $\sim$ 0.05 pixels,
or 0\farcs005.
The image for each epoch is scaled to provide the number of \elec per pixel
detected in a 6000 second exposure.
The same grid of bright external galaxies used to provide registration is
used to establish a common intensity normalization (corrections at level
of $\sim$1\%) for the second epoch relative to the first.
With normalized and well registered images for the two epochs a simple
difference image provides a sensitive means of detecting objects that vary
in position (some nearby stars) or intensity (AGNs and Supernovae, see GNP
for the latter).
As an example of the exquisite sensitivity provided by the data, GNP show
that 90\% of the SNe at $m_{AB(8000)}$ = 26.4 in either epoch associated with
z $\sim$ 1.2 galaxies could be securely detected, with the completeness falling
to 50\% at $m_{AB}$ = 27.8.
For the purpose of detecting change with time the extreme stability provided
by $HST$ is an important factor in being able to probe to unique magnitude
depths with these observations.

\subsection{Sensitivity to Photometric Changes for Galaxies}

In principle establishing the photometric change for any galaxy is as simple
as defining an appropriate aperture and comparing counts in the difference
image compared to the counts in the same aperture for the direct image.
In the next subsection we will see that the important additional subtlety
of changes over time for Charge Transfer Efficiency (CTE) must be dealt with.
In the remainder of this subsection we deal with simpler issues of galaxy
selection, aperture definition and quantification of expected errors.

We adopt the compilation of 1067 galaxies from Fern\'{a}ndez-Soto, Lanzetta
and Yahil (1999, hereafter: FLY) for the HDF-N.
This is an I-band (F814W) magnitude limited survey
comprised of two sets of galaxies from two defined
zones in the HDF-N based on the different apparent magnitude depths
achievable over the WFPC2 CCDs.  The first zone covers the majority
of the 3 Wide Field CCDs (3.92 arcmin$^2$) and contains 946 galaxies
complete to m$_{AB(8140)}$=28.0.  
The second zone covers the Planetary Camera CCD and
the outer edges of the Wide Field CCDs (1.39 arcmin$^2$) and contains
121 galaxies complete to m$_{AB(8140)}$=26.0.

Each of these galaxies has a photometric redshift based on 7 color ($UBVIJHK$)
photometry-redshift relations, and where available a spectroscopic redshift
from the literature.
The photometric redshifts show an $rms$ dispersion of $\sigma$ = 0.09 in 
comparison
with a sample of $\sim$100 spectroscopic redshifts at z $<$ 2, with no cases
discordant at $>$ 3$\sigma$.
We have derived a coordinate transform between the $x, y$ positions as
tabulated by FLY and our factor of 4 over-sampled images;
this is accurate to $\lesssim$ 10 of our pixels (or 0\farcs25).
We then adopt the pixel with maximum counts within $\pm$ 20 over-sampled pixels
as the nominal center of each galaxy.

Williams \etal (1996) tabulate a first moment radius for each galaxy.
For an $r^{1/4}$ light distribution the half-light radius is one-half the
first moment radius, and we have simply adopted one-half the first moment
radius for each galaxy as a nominal scale for an aperture on each galaxy.
Figure 1 shows the distribution of galaxy half-light radii as a function of
apparent magnitude for the FLY sample.  

With a minimum aperture size of 0.05$\arcsec$, the aperture photometry
for some of the galaxies
essentially returns the central pixel value
in each epoch.  There is no reason to declare this a suspect value.
First, the typical variation in the intra-pixel QE response is
1 to 2\%, depending upon where a point source is centered within
the pixel.  This is reduced in amplitude 
by the square root of the number of dithers to different sub-pixel
coordinates resulting in an error of $\lesssim$0.5\%.  
At these magnitudes, the intrinsic random 
errors are already ten times this value.
Second, time dependent gain variations over the entire chip
are calibrated out in
establishing differential photometry with a mean of zero across epochs.  
Furthermore, time dependent gain variations for individual pixels, however
unlikely, would be eliminated when producing
our combined images.

At $m_{AB(8000)}$ = 26.0 an integrated object count level of $\sim$ 310 \elec
is accumulated in 1750 seconds (average exposure time of second epoch data).
At a half-light radius of 0\farcs2 half of this light would be spread over
4$\pi$ pixels (0\farcs1, WF), or $\sim$ 12.3 \elec per pixel.
The sky level per exposure is about 78 \elec per pixel on average.
If we make the reasonable estimate that the local sky can be defined to
$\sim$ 1 \elec per pixel, then with an aperture of 0\farcs2 at $m$ = 26
the uncertainty from possible zero point errors on the sky is 8\%.
The intrinsic Poisson, sky and readout 
noise limited S/N (within the half-light
radius) in this example would be $\sim$ 24 in the shallower 2nd epoch and
$\sim$ 34 in the original HDF.
At an uncertainty of 1 \elec in the zero point of the difference image,
photometric errors from this term alone are 60\% larger than the intrinsic
S/N limit (5\%, 1$\sigma$) of the difference image.
In order to not have uncertainties in the sky zero point dominate the errors
we have adopted aperture radii which are the minimum of tabulated values
and the quadratic relation plotted in Figure 1.
Adoption of this smaller radius, which is set at a point between the lower
envelope and the mid-point of measured radii, effectively removes uncertainties
in the sky zero point from major importance in an error budget.

Having defined an aperture choice the fractional change in photometric intensity
for any source is just the summed counts over this aperture in the '97 - '95
difference image divided by the summed counts in a grand average image.
In order to search for variable galaxies we now need to establish the
expected noise level of this measure for non-variable objects.
GNP developed two sets of ``null experiments" to help quantify the level
beyond which random noise could be misinterpreted as Supernovae.  The
control images that result from these null experiments are also ideal for 
establishing the expected galaxy-photometry
measurement noise as a function of magnitude.

One null experiment consists of dividing the full set of individual exposures
into two sets, one composed simply of the odd numbered frames and the other
the even numbered ones.
Each set is now an ``epoch", but with no real time difference.
The difference image from these even and odd epoch combined frames effectively
carries through the effects of object and sky Poisson noise, readout noise,
and any limitations imposed by minor errors in the sky zero points used.
However, no intrinsic variations (on timescales greater than exposure times of
individual frames) can exist and the apparent variations in photometry
between the sets can be used to define the noise-magnitude relation.

As a second null experiment,
GNP had also defined simulated images for each epoch starting with the same
over-sampled and optimally combined representation of the source distribution
and non-variable background light (noiseless relative to levels in 
individual exposures).  These combined images are represented
as bi-cubic surface fits to the light distribution over individual
WFPC2 pixels as a function of sub-pixel x and y offsets.  As such these
can be evaluated at the offset position of any individual exposure
and an amount of noise appropriate to readout, dark current, sky
background and source Poisson statistics added to simulate each
individual exposure.
The simulated frames for each epoch were then recombined to produce an
over-sampled image for each epoch.
Again, this should capture most sources of noise, but no real variations
can exist.

Figure 2 shows the adopted 1$\sigma$ galaxy photometry error as a function
of magnitude.  This represents a sum over results for the
two independent null experiments for which the differences between
the two were small.  The apertures used here and throughout the paper
are as shown in Figure 1, namely for each galaxy the minimum value
of the tabulated half-light radius and the quadratic function of magnitude.
The magnitudes adopted from here on (referred to as $m_{AB,int}$ later) apply
to the intensity level internal to the adopted half-light radii.
An appropriate zero point offset is adopted to bring the mean of these
generally smaller aperture magnitudes into registration with the FLY
isophotal magnitudes.
The data points represent the $rms$ photometry differences for all galaxies
within $\pm$ 0.45 magnitude intervals at $m$ = 22 with this interval
decreasing to $\pm$ 0.25 mag at $m$ = 27 from the two sets of null experiments.
Note that the simple first principles argument developed earlier gave an error
of 5\% at $m$ = 26 in excellent agreement with this calibration.
The distribution of errors in any magnitude bin for the null experiments
is well centered on zero and has a Gaussian distribution.
A lower limit to the assumed photometric error of 0.012 magnitudes was
adopted based on general experience with absolute limits to WFPC2 photometric
accuracy.
Our conclusions are not sensitive to reasonable bounds on this limiting
precision.

A direct solution for photometric changes of real galaxies, normalized
to the typical expected error as a function of magnitude can now be easily 
evaluated.  Figure 3 shows the distribution of photometric changes both as
direct delta intensities across the two epochs and after normalizing
these to the expected errors as a function of magnitude.
The results showed a very non-Gaussian distribution with 33 galaxies out of the
brightest 633 (selected on the basis of m$_{AB(8140)}$$<$27.5 for the standard
apertures as shown in Figure 1) 
showing $>$3$\sigma$ variation compared to statistically
expecting 2 of 633 by chance for a normal distribution.
Either the rate of variable galaxies in the HDF far exceeds the canonical
fraction of $\sim$2\% for AGNs, or we have unaccounted for noise or 
systematic errors.  Inspection of the photometric differences across 
epochs showed clear correlations
with magnitude and with $y$-position on the CCDs.
Such behavior is suggestive of a CTE systematic error term.
In our case, since we are comparing data acquired with similar exposure times
and nearly identical pointings, such an effect would follow only from a 
change of CTE with time.

\subsection{The Role of Charge Transfer Efficiency Changes}

We have solved empirically for corrections to the photometry with a simple
linear model of time differential CTE having linear terms in magnitude
and source position on each CCD.
An ideal solution will eliminate any systematics in photometry differences
between epochs.
Experimentation showed that any effect in $x$ must be significantly smaller
than for $y$, and not finding any consistency across CCDs the $x$-dependence
for time-differential CTE was taken to be zero.
Generally consistent values for a CTE term with linear dependence on $y$
were found across the CCDs, and with restricting the solutions to either bright
($m$ =
24.5 $\pm$ 1.0) or faint ($m$ = 26.5 $\pm$ 0.5) galaxy sets.
The full range scatter for this term is $\pm$ 20\% evaluated separately for
the 3 WF CCDs and the $m$ = 24.5 set of galaxies; we have adopted the
mean values independent of CCD.
The linear correlation coefficients of relative intensity differences against $y
$-position
ranged from 0.23 to 0.32 (correlations against $x$ spanned -0.09 to 0.02).
The adopted CTE (time difference '97 - '95 epochs) is:

\begin{equation}
C \, = \, -0.029 \,+\, 0.006 * (24.5 - m_{AB}) \,-\, 1.0 \times 10^{-5} * (y - 1
600).
\end{equation}

\noindent
The sign of each term is consistent with a slightly increased CTE loss over the
two years between the HDF epochs.
The magnitude of the slope in $y$ produces a 3.2\% effect bottom to top. 
Since
for galaxies at $m_{AB} \, \sim$ 24.5 the assumed intrinsic noise is $\sim$ 
2.4\%,
such an overall slope is a significant bias.
Figure 4 shows the solution for a $y$-dependent correction factor for WF4 for
which the linear correlation coefficient is -0.29.

Our solution for a CTE effect is in a unique range of parameter space compared
to previous calibrations.
In particular our correction primarily applies to objects that have a 
comparable,
or smaller per pixel count level than local sky in individual exposures.
Published calibrations for CTE (Whitmore 1998, Stetson 1998) have generally 
relied on
well exposed stars (at least by the HDF standards) for which a typical star
image would peak at $>$ $10^3$ photoelectrons per pixel with an ambient diffuse
sky brightness of $\sim$ 10 \elec per pixel.
Even for ``bright" galaxies in the HDF with $m_{AB}$ = 24.5 and a median
half-light radius of 0\farcs15 the average count level for pixels within
this radius is 87 \elec compared to typical sky of 78 \elec.
Whitmore (1998) found an increased loss attributed to a change of CTE
for faint stars with $\sim$ 40 \elec per pixel (average within a 0\farcs2 radius
aperture) equal to 9.5 $\pm$ 3\% (center of CCD) over a 3 year baseline.
The HDF galaxies have smaller to comparable count levels, but are relative
to a higher sky level (which will suppress the loss) compared to the
calibration described by Whitmore (1998).
Stetson (1998) found a much smaller time dependence of $\sim$ 1\% over
3.5 years analyzing a more extensive set of data.

We note that a small relative component (10\% of total) of our derived
time dependent CTE change can simply be attributed (using the Stetson 1998
equations) to our use of shorter exposure times on average for the second
epoch (1750 s) compared to the original F814W HDF exposures (2130 s).
Our derivation of the time-differential CTE effect is well defined 
empirically for the
two epochs of HDF galaxy photometry, and given the suppression from a
larger ambient sky background is generally consistent with the changes found by
Whitmore (1998).  The Whitmore analyses are done differentially
comparing WFPC2 photometry acquired after large scale changes of
stellar position on the CCDs; the Stetson analyses for CTE directly
adopt a comparison with ground based, crowed-field photometry.
Published time dependent CTE results at significant amplitude only apply to the
very faintest
stars, for these, ground based systematic errors may not be negligible
and may explain the different level of effect found in the studies.
In any case the several hundred galaxies available in the HDF,
$\sim$ 99\% of which are non-variable, have been used here to
provide an empirical correction for time-dependent CTE changes
{\em that for these two epochs of F814W data} is reliable and
accurate for our purposes.

\subsection{Photometric Changes of HDF Galaxies}

With a solution for time-differential CTE included, any remaining systematic
errors in galaxy photometry across epochs will be small compared to random
errors per galaxy.
Figure 5 shows the full set of corrected galaxy intensity differences between
the two epochs.
Details for all galaxies with changes relative to the expected error (Figure 2
relation) larger than 3$\sigma$ are given in Table 1, with column definitions 
of:
(1) Galaxy number from Williams \etal 1996.
(2) Isophotal F814W magnitude in AB system from FLY.
(3) $m_{AB}$ internal to the adopted half-light radius.
(4) Redshift from FLY based on spectroscopic determination
if available (s) or photometric (p).
(5) Fractional change in intensity across epochs within $r_{1/2}$;
positive implies brighter in 2nd epoch.
(6) Significance of change obtained by ratioing to the expected
error as a function of magnitude (Figure 2).
(7) Half-light radius as minimum of one-half the Williams \etal
(1996) first-moment radius and the quadratic function plotted in Figure 1.
(8) Separation between centroid of excess light and the galactic
nucleus.
(9) Absolute value of the magnitude difference between the two
epochs.
Results are tabulated only for the 633 galaxies which have $m_{AB,int}$ (column
3 of Table 1) brighter than 27.5.

Figure 6 shows the cumulative distribution of the variability
significance (the ratio of the change in magnitude to the expected
error at that magnitude) for all of the galaxies considered in the
study.  
This value is given in column 6 of Table 1 for the 9 variable galaxies. 
We show the cumulative distribution (solid line), the fit
to these points at $\delta$/$\sigma$ $<$ 2.7 (dashed line), and the 
positions of the 9 variable
galaxies along the x-axis.  A clear excess of galaxies can be seen beyond 
$\delta$/$\sigma$ $=$ 2.7.  The area under the
dashed line beyond $\delta$/$\sigma$ $=$ 2.7 translates to               
approximately 2 galaxies, the number that would be expected in a normal 
distribution to have variability significance greater than 3$\sigma$.  
From this figure, it is clear
that there is an excess of objects having a variability significance of greater
than 3 (we have set the selection threshold
at the traditional 3$\sigma$ level, slightly higher than the break at 2.7).

Of the nine variable galaxies listed in Table 1, 4-403.0 has been thoroughly
discussed in GNP.
The brightening in 4-403.0 has been attributed to a Supernova (SN1997ff) 
detected
in the second epoch data.
The residual light associated with 4-403.0 from subtraction of the first epoch
from the second is well represented as a point source 0\farcs16 off the galaxy
nucleus with $m_{AB}$ = 27.4.
This brightness is consistent with a Type Ia SN at the galaxy redshift of z = 
1.32
detected $\sim$ 40 days (observer's frame) after maximum.
This off-center brightening was judged much more consistent with a SNe 
interpretation
than as AGN variability.
A second SN (SN1997fg) associated with galaxy 3-221.0 was also reported in GNP;
it was not picked up in this galaxy variability search since the SN was well 
outside
the half-light radius summed over for 3-221.0.

The Supernova search used a different approach of detecting point sources 
present
in a difference image of the two epochs independent of knowledge of galaxy 
positions
(except for scaling errors up appropriately to account for increased Poisson 
noise
based on count levels).
The variable sources 2-251.0 and 3-404.2 were also briefly discussed in GNP.
In these cases the brightening comes from an apparent point source coincident
with the galactic nuclei within small measurement errors.
For 3-404.2 with a half-light radius of 0\farcs15 the formal offset of the
first epoch excess light was 0\farcs048 from the galaxy center (consistent with
zero)
and the amount of galaxy light within a radius of 0\farcs05 is $\sim$ 4\% of the
total
making interpretation as an AGN more likely than as a SN.
For 2-251.0 there is abundant supporting evidence in the literature (see \S 3)
that this is an AGN.

The next to last column of Table 1 gives the apparent radial offset ($\delta r$)
of the
excess light in one epoch from the galaxy nucleus.
The error on this measurement is typically about 0\farcs05.
In all cases, except for 4-403.0 for which we believe the brightening in the
second epoch was caused by SN1997ff, the estimated $\delta r$ is well under the
estimated half-light radius.
Coincidence with the nucleus alone cannot establish whether a specific event
should be attributed to AGN activity or a SN, but statistically the SNe would be
expected outside the $r_{1/2}$ radius as often as inside.
The strong bias for the excess light centroids in Table 1 to align with the
galactic nucleus to a level well under the half-light radius supports 
interpretation of these events as AGN.

\subsection{Additional Photometric Change Measures for 2-251.0}

From inspection of Table 1 it is clear that 2-251.0 is in a class by itself.
The absolute magnitude of the excess light across the epochs at $m_{AB} \, \sim$
23.5 is a factor of 20 brighter than the next brightest at $\sim$ 26.7.
Furthermore, even with the comparison of 63000 seconds in F814W with the 
original
longer HDF, 27th magnitude is where errors are becoming large.
From inspection of the two epochs of F300W data (2nd epoch only 15\% as long as
the original) it is obvious that: (1) 2-251.0 is clearly visible and exhibits
change.  (2) None of the other AGN/SNe candidates are even close to detection
in F300W.

We searched the Hubble Archive for any other pairings of HDF exposures giving
temporal information on the brightness of 2-251.0.
The pre-HDF test of guide stars six months before the original HDF included
three F606W exposures with a total exposure time of 2200 seconds with 2-251.0
present on WF3 at $x,y$ = 252,410.

Table 2 summarizes the available photometry for 2-251.0.
Over six months just prior to the HDF the F606W observations are consistent
with no change, or a slight increase of flux.
Over two years after the original HDF galaxy 2-251.0 brightened by
$\sim$ 8\% in F814W (rest wavelength $\sim$ 4100 $\pm$ 250 \AA ) and by $\sim$
40\% in F300W (rest wavelength $\sim$ 1500 $\pm$ 200 \AA ).

\section{The Variable Galaxies}

Figure 7 is a composite of the I-band images of the 
eight variable galaxies that are candidate
AGNs from Table 1.  These galaxies meet the criterion for having
nuclear variations in excess of 3$\sigma$ above the background noise.
However, we note that two such galaxies would be expected
to appear as 3$\sigma$ variables statistically in a normal distribution
as discussed in \S 2.4.
Each pair of images consists of the I-band image of the galaxy on the left
and the difference image of the two epochs 
shown to the right (first epoch
subtracted from the second epoch observation).  The gray scale has been
adjusted so that the variable region is always positive (black).
The dash at the bottom right in each frame
represents the 1kpc scale for the galaxy.  The galaxies appear to cover a
range of Hubble types and magnitudes.
We briefly discuss each of the galaxies below.

2-251.0 - This red galaxy at z = 0.96 (spectroscopic) 
is the most variable galaxy in our survey with a magnitude excess of
23.5 mag over the 2 year period.
This galaxy also has an interesting morphology. 
Two-dimensional model fitting finds a prominent bulge component (Marleau \&
Simard 1998) although spiral-like structure is also seen out to 
$\sim$0.6$\arcsec$ 
from the nucleus.  The 
spectrum for this object reveals broad MgII (2800$\AA$) emission (Phillips
\etal 1997; Songaila 1997) which is further evidence that this object contains
an AGN.  In addition this object is also a radio source having been
detected in the VLA survey of the HDF-N (Richards \etal 1998; Fomalont \etal 
1997).  
The VLA observations also
report variability of this source by about 30\% over an 18 month period.
The radio spectral index is $\alpha$ = -1.0$\pm$0.1 further suggesting
the AGN nature of this object.  This galaxy was also detected by ISO having
S$_{7\micron}$=52$\mu$Jy (Rowan-Robinson \etal 1997).  Although
they suggest this source is a massive starburst with a SFR of 200 
M$_{\odot}$/yr,
the corroborating evidence seems to suggest that the presence of an 
AGN is the more likely explanation. 

4-260.111 - This galaxy has a spectroscopic redshift of 0.96 and appears
to be a late-type spiral galaxy with 2-dimensional model fitting indicating
little or no bulge component.  
The variability is off-nucleus by 0.1$\arcsec$
although this is well within the half-light radius of the galaxy.  This object
is also an ISO source with S$_{7\micron}$=51$\mu$Jy (Rowan-Robinson \etal 1997).

2-210.0 - This galaxy with a spectroscopic redshift of 0.75 also appears to
be a spiral with little or no bulge component.  The disk is ``blobby"
appearing to have several HII-like regions in the spiral arms.  The peak in
the variability is very close (0.03$\arcsec$) to the nucleus of this galaxy.

3-266.0 - Two-dimensional model fitting of this galaxy indicates the presence
of a large bulge component.  The small half-light radius (0.1$\arcsec$) is 
further evidence that this galaxy is elliptical.  A spectrum has just
recently been obtained for this object (Cohen et al. 2000) and the
redshift is 0.95. 

3-404.2 - This galaxy has a spectroscopic redshift of 0.52.  It has
a moderate bulge component (B/T=0.3 from Marleau \& Simard 1998)
and is probably an Sa-type spiral galaxy.

2-787.0 - This galaxy has a photometric redshift of 0.92.  The model
fitting indicates a small bulge fraction (B/T=0.08).  This galaxy,
in addition to the last two, are the least variable galaxies to meet
the selection criterion.  

4-327.0 and 4-344.0 - These two galaxies are very faint making their
morphologies difficult to classify.  Their photometric redshifts place
them at 1.9 and 1.4 respectively.  These are likely to be dwarf galaxies 
having absolute magnitudes around M$_I$$\simeq$-17.5.  

Although spectroscopic data exists for 5 of the AGN candidates selected in
this study, the spectra themselves or information regarding their spectral
properties were only available for object 2-251.0 at the time of
publication.

\section{Comparison with the Local Number Density of Seyferts}

One issue we would like to address with the results of this study is
the evolution of the number density of AGN.  Of the nine variable
galaxies detected, we are confident that one 
(4-403.0, see Gilliland, Nugent and Phillips 1999)
is a supernovae.
The majority of the remaining 8 are likely to be AGNs with the
caveat that two of these would be expected statistically from the 
noise distribution tail.  By comparing the number
density of these variable galaxies at $<$z$>$$\simeq$1 to the local number
density of Seyfert galaxies, we can place limits on any 
significant evolution of the population of AGNs over this
redshift range.

\subsection{The Luminosity Function of the Variable AGN}

To compare the number 
density of variable AGN detected within HDF-N galaxies to that of 
local Seyferts,
we require an estimate of the absolute magnitude of the 
AGN component in each of the variable galaxies.  
To make this estimation, we must approximate the expected level of variability
for each AGN.  Then, based upon the amount of variability observed
for each galaxy, we can estimate the fraction of the flux that is due
to the AGN.  It is difficult, however, to know the precise variability
amplitude expected for a given AGN.  
Many studies have shown that the amplitude of variability
depends on the AGN luminosity (Trevese \etal 1994, Hook \etal 1994).
Unfortunately, surveys
of low-luminosity AGN variability have not yet been completed and therefore,
little exists in the way of statistical results which quantify the amount
of variability expected for the active nuclei we are trying to detect
in this study.  At this point, the best we can do is estimate the amount
of variability expected for our nuclei based on the results of larger
variability surveys of intrinsically brighter QSOs and attempt to extrapolate
to lower luminosities based on the apparent trends in amplitude with magnitude.

We adopt the structure function results from Hook \etal (1994) 
to estimate the expected variability amplitude
for our AGN over a given time interval.  The structure function was
calculated from a sample of $\sim$300 optically selected quasars studied
over a period of 16 years at seven epochs.  From this sample of QSOs,
an amplitude variation with respect to absolute magnitude was detected.
The structure function is fit with the equation  
\begin{equation}
S=\sqrt{[0.155+0.023(M_B+25.7)]t^p}
\end{equation}
where M$_B$ is the absolute magnitude of the QSO.
When the observed time interval {\it t} is known, this
equation yields the expected amplitude of variability in magnitudes for a QSO. 
The value {\it t} in equation 2 is two years for an object at z=0
in the HDF images
used in this study.  However, higher redshift objects have decreasing
values of {\it t} due to time dilation effects.  To determine the
true observed time interval for each galaxy, the actual observational
time interval of 2 years must be divided by (1+z$_{gal}$) such that
a galaxy at z=1 would have {\it t}=1 year.  The redshifts we use are
a combination of spectroscopic redshifts when available and
photometric redshifts (FLY).

With equation 2, we estimate the expected amplitude of variability 
for each AGN and determine the fraction of the galaxy flux due to the AGN
component based on the observed variability of the galaxy.  In combination
with the known redshift, we then know the absolute magnitude of
the galaxy and AGN component.
The Hook \etal sample included only QSOs with M$_B$$\leq$-23.0 so to
apply these results to our data, we must extrapolate to lower luminosities.
Since the structure function equation used to determine the AGN magnitude
is also a function of that magnitude, we use an iterative method to arrive
at a magnitude estimate.  We start by assuming M$_B$=-23.0, determining the
AGN magnitude from the variation amplitude calculated from equation 2, and
then inputing this new AGN magnitude into the equation to redetermine the 
expected level of magnitude variation.  We continue this process until we
reach the AGN magnitude which satisfies the equation and the observed level
of magnitude variation.  In all cases, this point was reached within
three iterations.  The estimated absolute magnitude range for our AGN is
-19.5$\lesssim$M$_B$$\lesssim$-15.0.
The absolute magnitude of the AGN,
as estimated from the structure function, has an associated error 
of $\pm$$\sim$0.25 mag which falls well within the bin size used to
to construct the luminosity function below. 

The luminosity function is computed using
the 1/V$_a$ technique (Schmidt and Green 1983) where V$_a$ is the
accessible volume over which each object can be observed in the survey.  
For computing the LF, we 
consider only zone 1 of the HDF which comprises the majority of
the 3 Wide Field chips, covering 3.92 arcmin$^2$.  All of our AGN candidates
fall within this region of the CCD.  

From the total integrated magnitude of the galaxy, we determine the 
distance at which each object could be observed within the surveys' 
apparent magnitude limit. 
We then determine the distance where the variability
amplitude of the AGN within the host galaxy falls below the 3$\sigma$ 
variability selection threshold.  The accessible volume, V$_a$, is then
determined from the closer of these two distances.  
Table 3 lists the absolute magnitude for each variable nucleus 
included in the LF calculation as well as the apparent I magnitude,
the maximum redshift over which the object could be observed (z$_{max}$),
and V$_a$ for each object in Mpc$^3$. 
The LF is the summation
of 1/V$_a$ for each object divided into two absolute magnitude bins
centered on M$_B$=-15.75 and -18.25 (Figure 8).  The luminosity function
is tabulated in Table 4.

The determination of the apparent magnitude limit of the survey is
an important aspect of the LF calculation.  The number
counts of the HDF-N galaxies in FLY appear to turn over at $\sim$26.5 to
$\sim$27.  We also note that FLY indicate that the median redshift of
their galaxy catalog steadily increases up to I$_{AB}$=26.0 and then 
remains fairly constant.  To avoid incompleteness due to surface brightness
selection effects as well as possible redshift incompleteness we have
set the apparent magnitude limit for the calculation of the LF at
I$_{AB}$=26.0.  In doing this, we must remove the two faintest variable
sources, 4-327.0 and 4-344.0, from our LF calculation since their 
integrated galaxy magnitudes fall below this magnitude cut-off.  
These objects also have the highest estimated redshifts among our candidates 
(z$_{phot}$$>$1) as well as displaying some of the lowest variability
significance according to Figure 6.  

\subsection{Discussion of the Luminosity Function}

Figure 8 is the resulting
LF for the 6 brightest variable galaxies in the HDF-N.
We compare the LF for the variable galactic nuclei in 
the HDF-N (squares) 
with the local Seyfert LF (Seyfert 1s and 2s) from Huchra \& Burg 
(1992) (filled circles) and the QSO LF at $<$z$>$=1 from Hartwick \& Schade
(1990) (asterisks).
We have converted our I magnitudes for the variable HDF galaxies into B
magnitudes and applied a K-correction based upon the redshift assuming 
a power-law spectrum for the AGN candidates with $\alpha$=--1.0.
With these assumptions, the K-correction is zero and 
I$_{AB}$--B$_{Johnson}$=--0.64.
The error bars represent the Poisson errors based on the number
of objects in each bin.  Due to the small number of objects used in
constructing this LF, we have chosen a large bin size (2.5 magnitudes) 
to reduce the statistical error per bin.

The most remarkable feature of this figure is the smooth transition
between the local Seyfert LF and 
our AGN candidates from the HDF.  We note that our AGN LF extends to
fainter magnitudes than the local LF even though our data are at higher 
redshifts ($<$z$>$=0.84) demonstrating the strength of this detection
technique.  In the luminosity range where our LF overlaps the local
Seyfert LF, there is no evidence for an increase in number density.
Within the error bars, the number of AGN at -18.5$\geq$M$_B$$\geq$-19.5 
remains the same for the local and higher redshift LFs.
This is in stark contrast to the situation at brighter absolute magnitudes
(M$_B$$\lesssim$-23.0) where the number density appears to increase
significantly between local and higher redshift surveys.   

The shape of our LF, although comprised of only two luminosity bins, does 
display the trend of increasing number density with fainter
absolute magnitude.  The bin centered at M$_B$=-15.75 is higher than
that at -18.25 at greater than the 1$\sigma$ signficance level.   
Although clearly more AGN are necessary to verify this trend, 
this initial study points toward the possibility that AGN number 
counts have not turned over by M$_B$$\simeq$-16 but continue to increase
at these faint magnitudes.  More data are needed to determine the
behavior of the faint end of the AGN LF with greater statistical 
significance.

\section{Summary and Conclusions}

With the goal of investigating the number density of AGN at z$\simeq$1,
we have completed a survey of variable galaxies in the Hubble Deep Field North.
Through the detection of Seyfert-like galaxies based on their variable
properties, we
hope to help bridge the gap between luminous QSOs and intrinsically fainter
AGNs at higher redshifts.  This initial study demonstrates the power of
this technique for detecting Seyfert galaxies or low-luminosity AGN at
z$\sim$1.

Out of 633 galaxies with I$_{int}$$\leq$27.5, we have found 8 galaxies
which appear to be varying at $\geq$3$\sigma$ over the two year period.
These galaxies, at $<$z$>$$\sim$1, cover a range of 
absolute magnitudes, colors, and hubble types although the majority 
appear to have some
spiral structure.  At least one is a confirmed Seyfert 1 galaxy based
on spectroscopic and radio observations.

We have estimated the magnitudes for the variable nuclei found in HDF-N
galaxies by adopting a structure function for QSO variability. 
By estimating the expected variability amplitude for each candidate
AGN, we can disentangle the AGN flux from that of the underlying galaxy
and determine the magnitude of the varying nucleus.  We estimate the
nuclear magnitudes to lie in the range -19.5$\lesssim$M$_B$$\lesssim$-15.0. 

The comparison of the luminosity function for our variable nuclei to
that for local Seyferts does not indicate any number density
evolution for AGN at M$_B$$\simeq$-19.  
This variability selection technique, however, 
may result in an underestimate of the number of AGN due
to the fact that every AGN may not display variability
over the 2 year timescale we have studied.  Although it is difficult to know
exactly what fraction of AGN we have detected, we can make an estimate based
upon previous QSO variability studies.  The study
of SA57 (Trevese \etal 1994; Bershady \etal 1998) finds that
virtually all of the known quasars in this field were detected to be
variable.  They also point out that the fractional variability amplitude,
being higher for lower luminosity AGN, 
suggests an even higher success rate for
Seyfert-like galaxy detection, consistent with 
the results of Hook \etal (1994).  A recent
study by Giveon \etal (1999) confirms this anticorrelation between
variability amplitude and luminosity.  This study of 42 optically selected
PG quasars also finds that all were variable over the 7-year study period. 
Based upon the published amplitude gradients, 
we would estimate that $\sim$90\%
would have been detected in our survey over the 2-year period 
given our typical photometric errors.

Therefore, we estimate that the majority of Seyfert-like galaxies in the
HDF-N would be detected in our variability survey.    
With a conservative estimate that the detection rate is between 50 and 90\%, 
the total number of variable galaxies in the HDF-N at I$\lesssim$27
represents $\sim$1--3\% of all galaxies at
this magnitude limit.  Our result is consistent with the findings of 
Sarajedini \etal (1999) where the fraction of field galaxies
containing an AGN is estimated based on the presence of an 
unresolved nuclear component.
In this study, the fraction of galaxies containing a nuclear point-source
in HST imaged galaxies to z$\simeq$0.8 was determined as a function of the 
nucleus--to--host galaxy luminosity ratio.
The variable galaxies in the HDF-N are likely to contain AGN components that
comprise at least 15\% of the total galaxy light estimated from the observed 
level of variability in each selected galaxy.  Sarajedini \etal find that
$\sim$3\% of their field galaxies contain nuclear components which make
up $\geq$15\% of the galaxy's light, consistent with the results of this
study.  

As larger variability surveys of field galaxies are conducted with longer
temporal baselines, the number of low-luminosity AGNs detected will increase
allowing for more statistically complete studies of this population of
objects.  The results of this study demonstrate the strength and efficiency
of this technique to select low-luminosity AGN candidates at moderate
redshifts.

\acknowledgments

We would like to thank David Koo for many helpful suggestions and comments
on this paper.  We also thank the referee for carefully reading this paper
and making several suggestions to help improve it.
Support for this work was provided by NASA through grant
number AR-7984.02-96A from the Space Telescope Science Institute, which
is operated by the Association of Universities for Research in Astronomy,
Inc. under NASA contract NAS5-26555. 
V.L.S. completed part of this work while at the Astronomy 
and Astrophysics Department at UC Santa Cruz.  

R.L.G. acknowledges the hospitality of the Astronomy and Astrophysics
Department at UC Santa Cruz where much of this work was done, and the
ST ScI for a sabbatical leave, and support via GO-6473.01-95A and
AR-7984.01-96A.

%References

%Begin Figure Captions
\newpage

\centerline{FIGURE CAPTIONS}

\figcaption[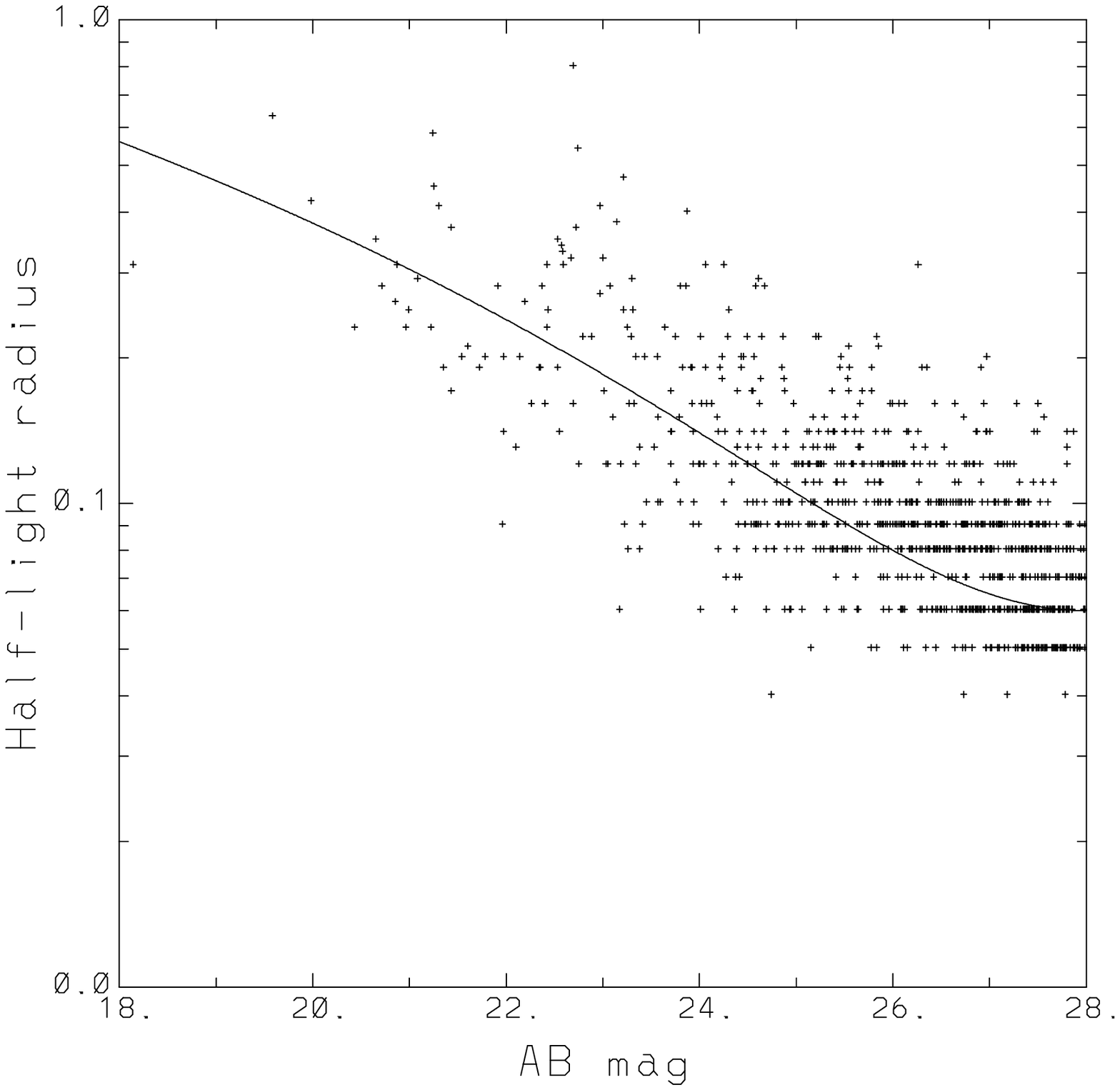]{
Half-light radii (half of first moment radii from Williams \etal 1996) 
in \arcsec plotted against isophotal F814W AB magnitudes from FLY.
The quadratic curve shows the adopted aperture radius for cases in which
this is less than the tabulated half-light radius.
\label{fig1}}

\figcaption[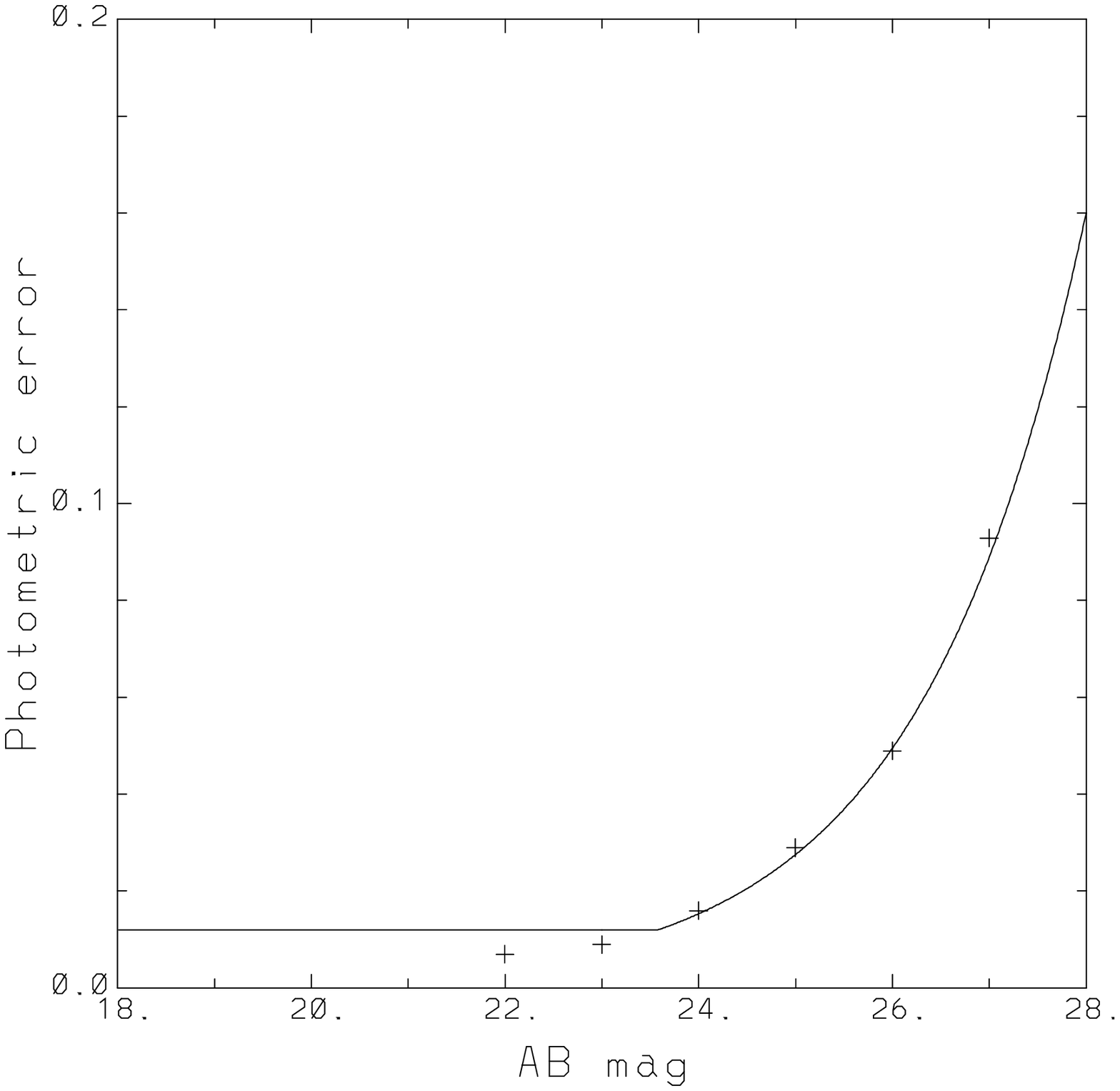]{
Plus signs show $rms$ of photometric relative intensity differences 
between galaxies in
the null experiments discussed in the text within given magnitude intervals.
The curve is a quadratic fit supplemented with a minimum value of 0.012.
\label{fig2}}

\figcaption[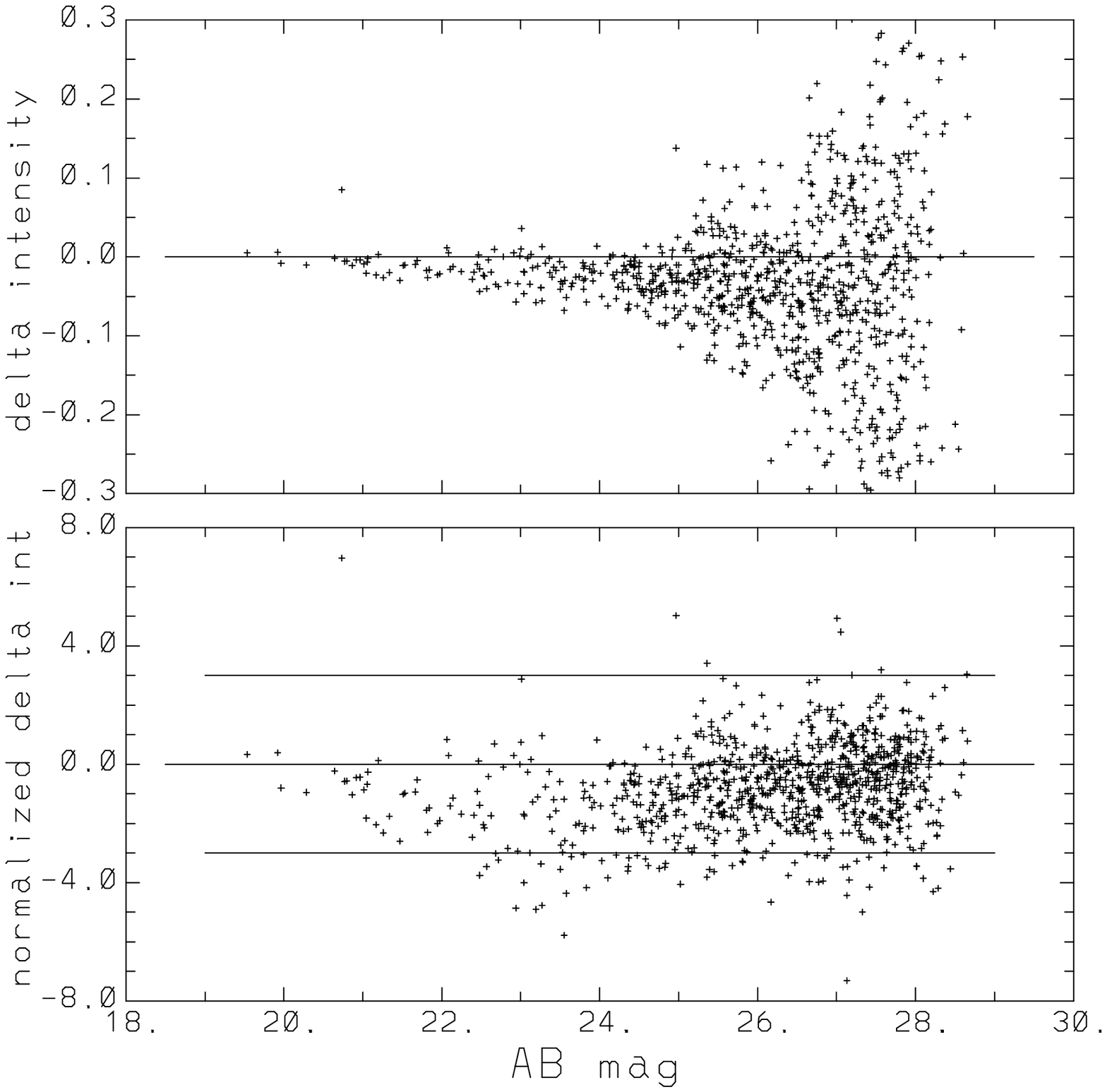]{
The upper panel shows the relative intensities of the HDF galaxies as the
difference of '97 minus '95 real data epochs relative to the grand average
data image intensity.
The lower panel shows the intensity differences after a magnitude
dependent normalization to the function plotted in Figure 2.
These are for intensity comparisons before application of the CTE
corrections discussed in \S 2.3.
Note the asymmetric distribution relative to the zero line, and the
large number of galaxies apparently variable at $>$ 3$\sigma$ levels.
The solid lines in the lower panel indicate the 0 and 3$\sigma$ levels. 
\label{fig3}}

\figcaption[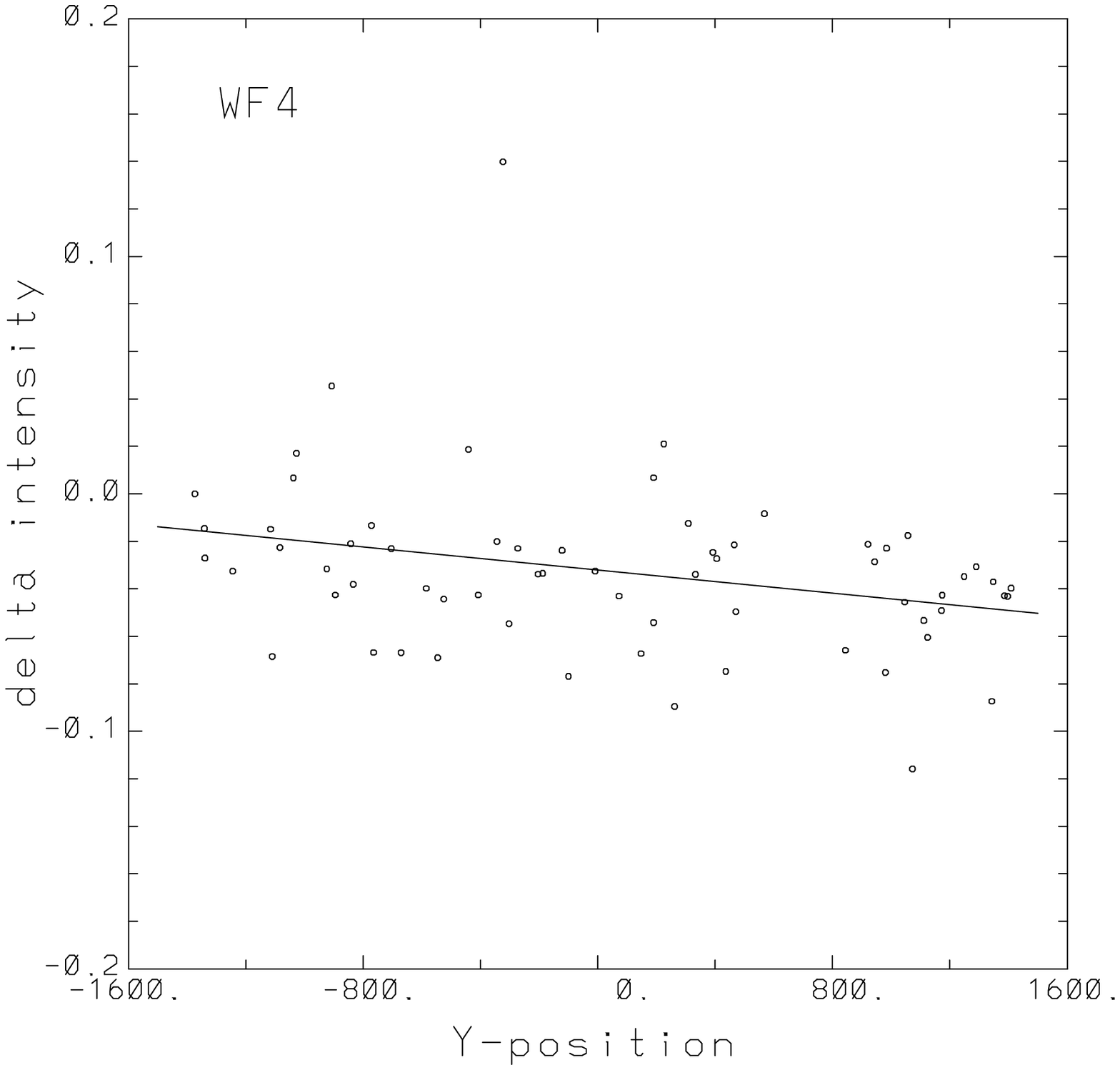]{
Demonstration of $y$-dependent CTE term showing the distribution of intensity
differences between the two epochs plotted against $y$ - 1600 (in our 
over-sampled by 4 data).
The $\sim$ 3.2\% CTE effect bottom to top of the WFPC2 CCDs (time-differential
across two years) introduces a strong photometry error if not corrected for.
\label{fig4}}

\figcaption[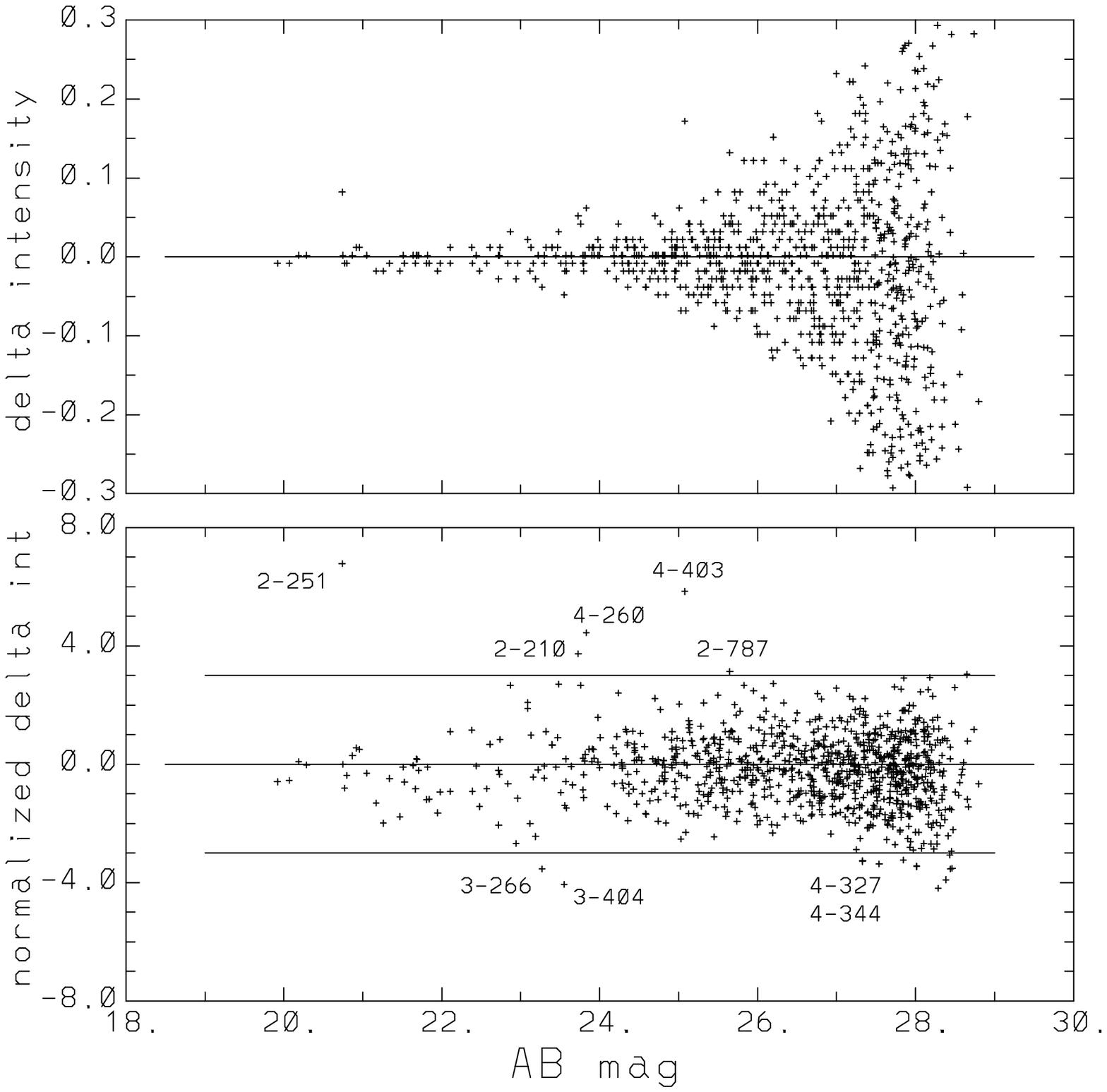]{
The intensity differences for the HDF galaxies after correction for the
time-dependent CTE are plotted in the upper panel.
The lower panel shows the same following normalization by expected random
errors per galaxy.  Again, the solid lines in the lower panel indicate 
the 0 and 3$\sigma$ levels.  The 9 galaxies varying by more than 3$\sigma$
at I(AB)$<$27.5 are labeled. 
\label{fig5}}

\figcaption[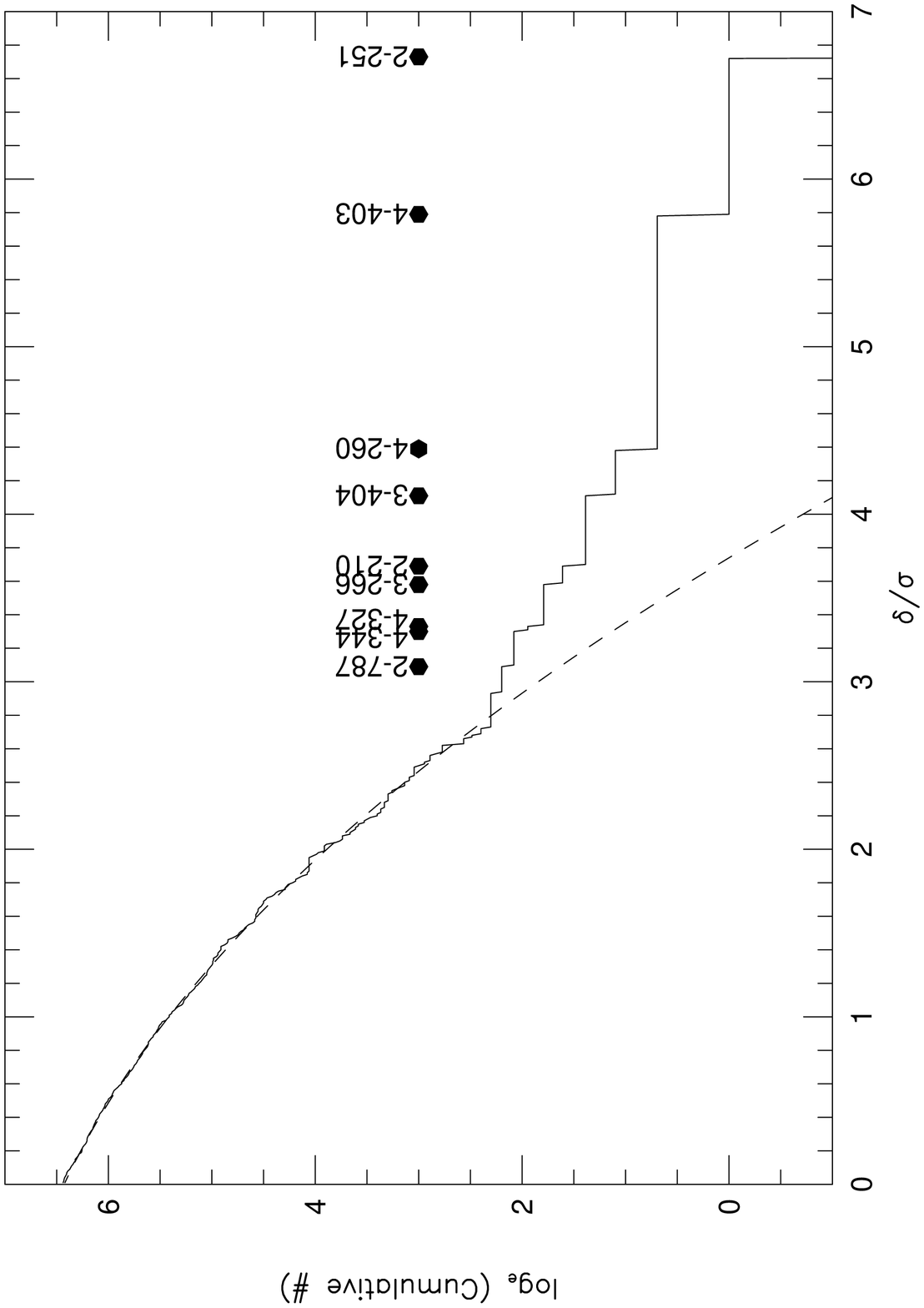]{
The cumulative distribution of the variability
significance (the ratio of the change in magnitude to the expected
error at that magnitude) for all of the HDF-N galaxies to 
m$_{AB(8140)}$=27.5 (solid line).
The fit to the data at $\delta$/$\sigma$ $<$ 2.7 is shown as the dashed 
line.  The positions of the 9 variable galaxies are marked along the x-axis.
\label{fig6}}

\figcaption[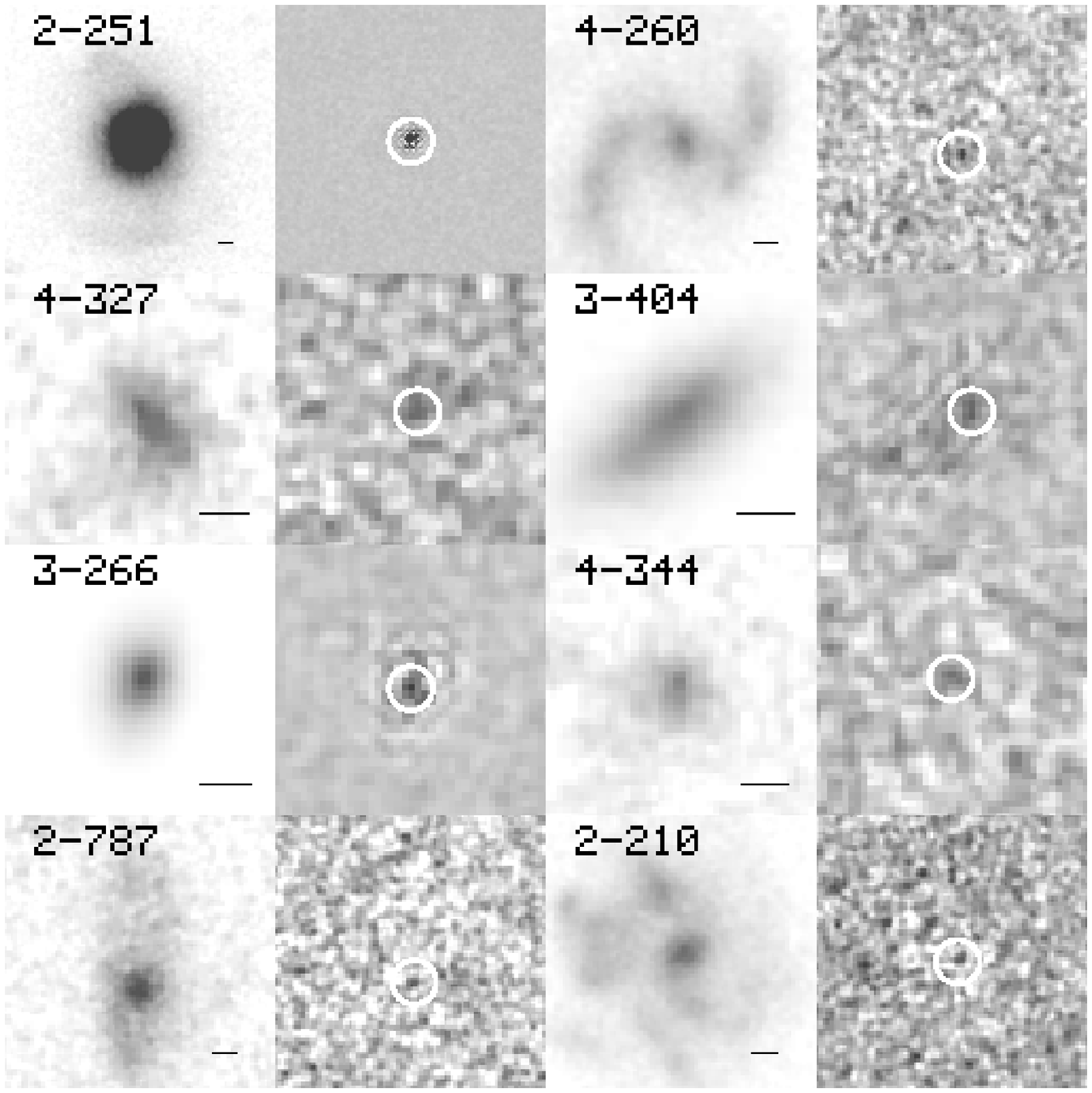]{
I$_{F814W}$ band image of the eight variable galaxies (left)
and the difference image between the two epochs (right).
The white circle on the difference image indicates the peak in variability.
The dash at the bottom right corner of each frame
represents the length of 1kpc in the rest-frame of each galaxy.
\label{fig7}}

\figcaption[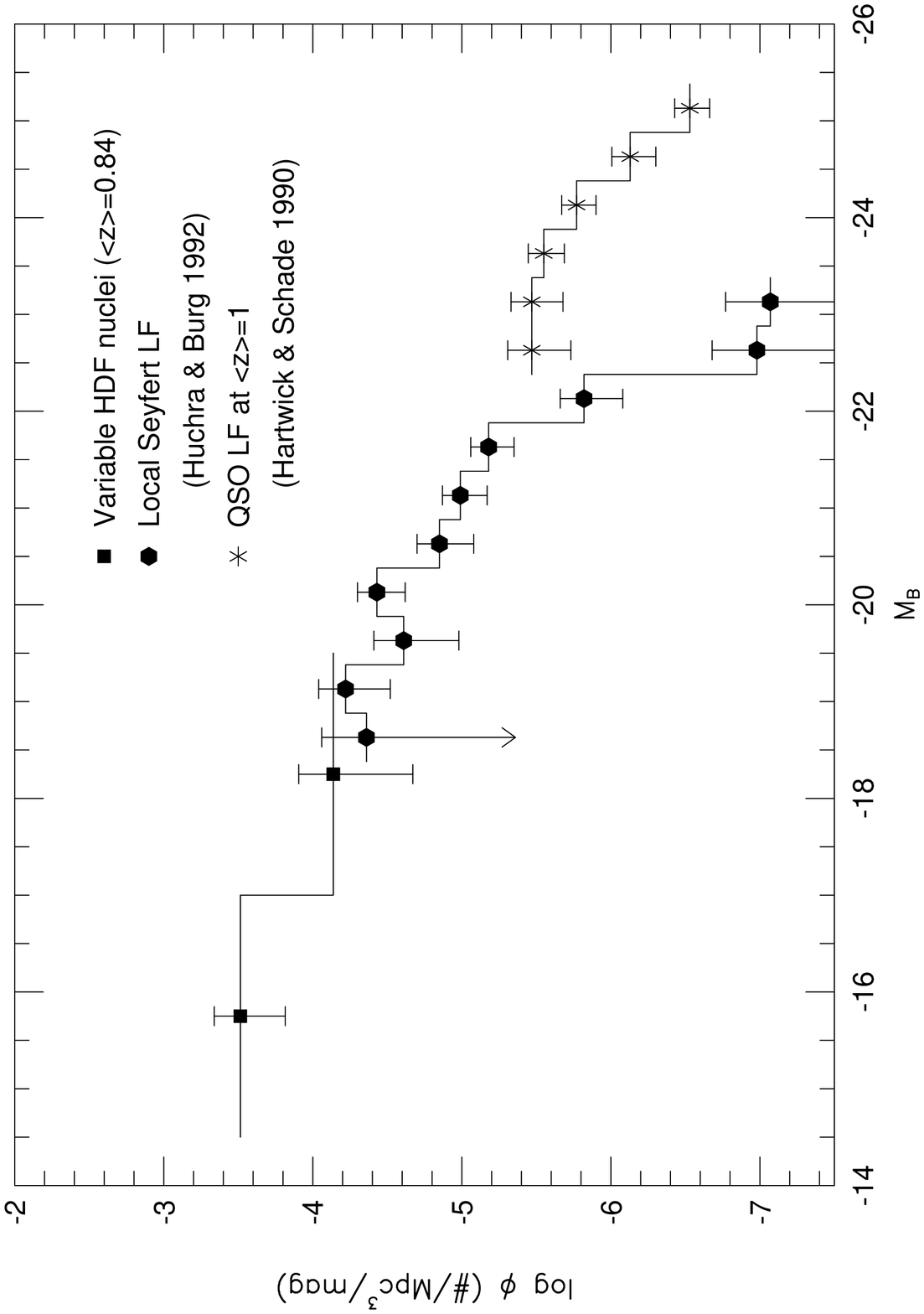]{
The luminosity function for our 6 AGN candidates having
integrated galaxy magnitudes brighter than m$_{AB(8140)}$=26
(filled squares). 
We compare our LF with the local combined Seyfert 1 and 2
LF of Huchra and Burg (1992) (filled circles) and the
QSO LF at $<$z$>$=1 (Hartwick and Schade 1990) (asterisks).
\label{fig8}}

%Begin Tables

\clearpage

\begin{center}
\begin{deluxetable}{lcccccccc}
\tablecaption{Variable Galaxies in the HDF}
\tablewidth{0pt}
\tablehead{
\colhead{ID} &
\colhead{$m_{AB}$} &
\colhead{$m_{AB,int}$} &
\colhead{z} &
\colhead{$\delta _{int}$} &
\colhead{$\delta _{int} / \sigma $} &
\colhead{$r_{1/2}$(\arcsec)} &
\colhead{$\delta r$(\arcsec)} &
\colhead{$m_{AB,excess}$}
\nl
\colhead{(1)} & \colhead{(2)} & \colhead{(3)} &
\colhead{(4)} & \colhead{(5)} & \colhead{(6)} & \colhead{(7)} &
\colhead{(8)} & \colhead{(9)}}

\startdata
2-251.0 & 21.24 & 20.76 & 0.96 s & 0.08 & 6.73 & 0.23 & 0.03 & 23.49 \nl
4-260.111 & 22.55 & 23.85 & 0.96 s & 0.06 & 4.39 & 0.21 & 0.11 & 26.88 \nl
2-210.0 & 22.69 & 23.75 & 0.75 s & 0.05 & 3.69 & 0.20 & 0.03 & 27.03 \nl
3-266.0 & 23.59 & 23.29 & 0.95 s & -0.04 & -3.58 & 0.10 & 0.06 & 26.71 \nl
3-404.2 & 23.59 & 23.57 & 0.52 s & -0.05 & -4.11 & 0.15 & 0.05 & 26.84 \nl
2-787.0 & 24.66 & 25.67 & 0.92 p & 0.13 & 3.09 & 0.12 & 0.06 & 27.92 \nl
4-403.0 & 24.78 & 25.10 & 1.32 p & 0.17 & 5.79 & 0.11 & 0.16 & 27.03 \nl
4-327.0 & 26.67 & 27.35 & 1.88 p & -0.36 & -3.33 & 0.07 & 0.03 & 28.45 \nl
4-344.0 & 26.93 & 27.35 & 1.40 p & -0.36 & -3.30 & 0.07 & 0.03 & 28.46 \nl

\enddata

\end{deluxetable}
\end{center}

\clearpage

\begin{center}
\begin{deluxetable}{cccc}
\tablecaption{Photometric Changes of 2-251.0}
\tablewidth{0pt}
\tablehead{
\colhead{Bandpass} &
\colhead{$m_{AB}$\tablenotemark{a}} &
\colhead{$\delta$ epoch (days)\tablenotemark{b}} &
\colhead{$\delta m$\tablenotemark{c}}}

\startdata
F300W & 25.32 & +732 & -0.37 $\pm$ 0.05 \nl
F606W & 22.52 & -190 & -0.02 $\pm$ 0.03 \nl
F814W & 21.35 & +732 & -0.088 $\pm$ 0.012 \nl

\tablenotetext{a}{Isophotal magnitudes in AB system from
Williams \etal 1996.}
\tablenotetext{b}{Difference between epochs in days relative to the primary
HDF observations.}
\tablenotetext{c}{Difference in magnitudes always in the sense of later
epoch relative to the earlier.}

\enddata
\end{deluxetable}
\end{center}

\clearpage
\begin{center}
\begin{deluxetable}{ccccc}
\tablecaption{AGN Component of HDF Variable Galaxies}
\tablewidth{0pt}
\tablehead{
\colhead{ID} &
\colhead{m$_{AB(8140)}$} &
\colhead{M$_B$} &
\colhead{z$_{max}$} &
\colhead{V$_a$}}

\startdata
4-260 & 22.55 & -17.6648 & 2.11 &  4596.81 \nl
3-404 & 23.59 & -14.8284 & 0.66 &  632.229 \nl
3-266 & 23.59 & -16.1090 & 1.02 & 1479.11 \nl
2-787 & 24.66 & -16.2056 & 1.46 & 2697.36 \nl
2-210 & 22.69 & -16.7188 & 1.38 & 2453.97 \nl
2-251 & 21.24 & -19.3116 & 4.41 & 10516.23 \nl

\enddata
\end{deluxetable}
\end{center}

\begin{center}
\begin{deluxetable}{ccc}
\tablecaption{Luminosity Function for HDF Variable Galaxies}
\tablewidth{0pt}
\tablehead{
\colhead{M$_B$} &
\colhead{$\phi$} &
\colhead{n}}

\startdata
-15.75 & -3.51 & 4 \nl
-18.25 & -4.13 & 2 \nl 

\enddata
\end{deluxetable}
\end{center}

\end{document}